\documentclass[aps,pre,twocolumn]{revtex4-1}

\usepackage{amsmath}
\usepackage{amssymb}
\usepackage{color}
\usepackage{subfigure}

\usepackage{graphicx}

\DeclareMathAlphabet{\mathitbf}{OML}{cmm}{b}{it}

\newcommand{\sFrac}[2]{{\textstyle\frac{#1}{#2}}}
\newcommand{\tripleCdot}{\stackrel{\mbox{\bf\scriptsize .}}{:}}

\begin{document}
\title{Nonlinear plastic modes in disordered solids}
\author{ Luka Gartner and Edan Lerner}
\affiliation{Institute for Theoretical Physics, Institute of Physics, University of Amsterdam,Science Park 904, 1098 XH Amsterdam, The Netherlands}

\begin{abstract}

We propose a theoretical framework within which a robust \emph{mechanical} definition of precursors to plastic instabilities, often termed `soft-spots', naturally emerges. They are shown to be collective displacements (modes) $\hat{z}$ that correspond to local minima of a `barrier function' $b(\hat{z})$. The latter is derived from the cubic expansion of the variation $\delta U_{\hat{z}}(s)$ of the potential energy upon displacing particles a distance $s$ along $\hat{z}$. We show that modes corresponding to low-lying minima of $b(\hat{z})$ lead to transitions over energy barriers in the glass, and are therefore associated with highly asymmetric variations $\delta U_{\hat{z}}(s)$ with $s$. We further demonstrate how a heuristic search for local minima of $b(\hat{z})$ can a-priori detect the locus and geometry of imminent plastic instabilities with remarkable accuracy, at strains as large as $\gamma_c-\gamma \sim 10^{-2}$ away from the instability strain $\gamma_c$. Our findings suggest that the a-priori detection of the soft-spots field in model glasses can be effectively carried out by the investigation of the landscape of $b(\hat{z})$.

\end{abstract}

\maketitle

\section{introduction}

\begin{figure}[!ht]
\centering
\includegraphics[scale = 0.57]{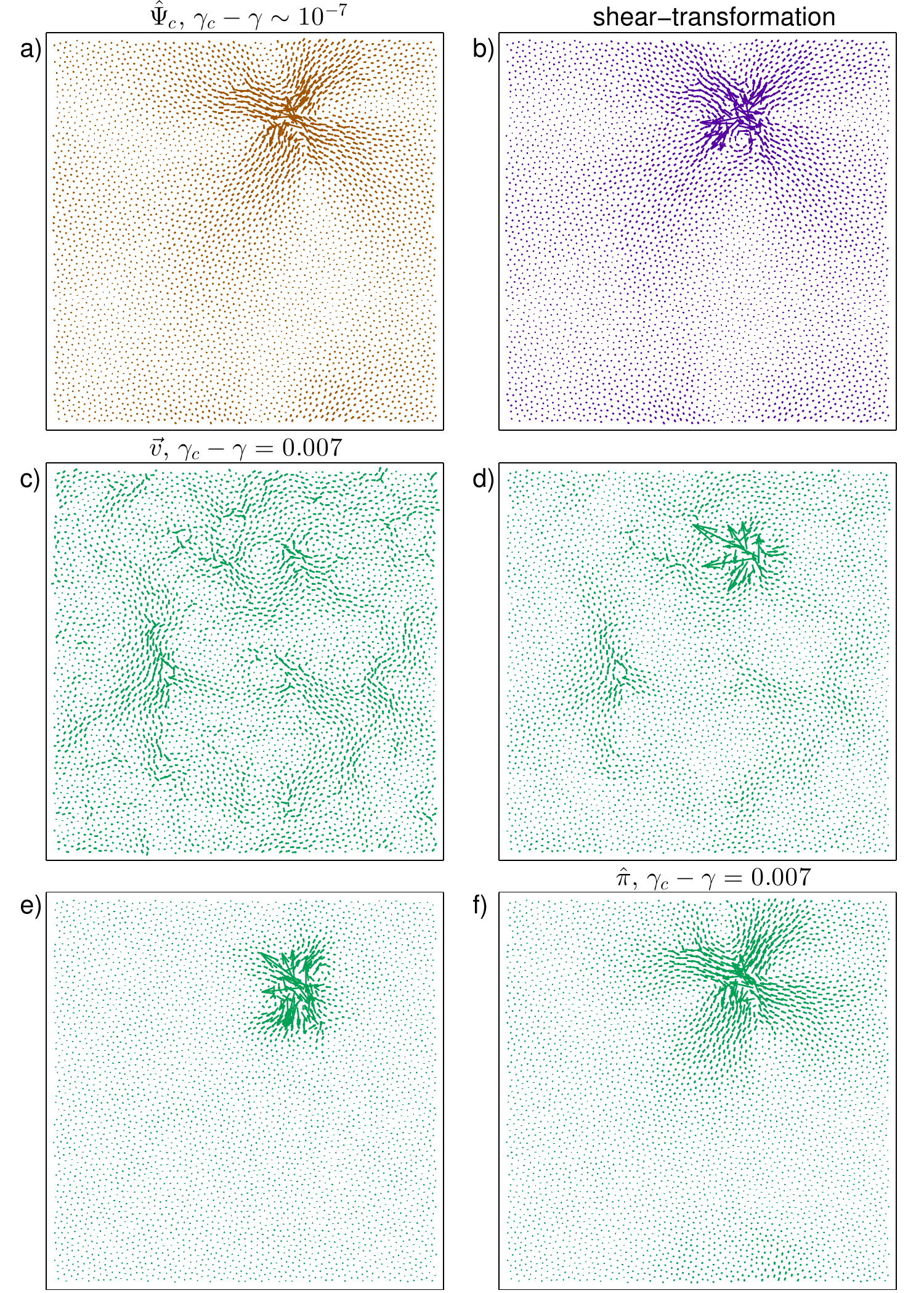}
\caption{\footnotesize (color online). A plastic instability in a sheared two-dimensional model glass. {\bf a)} The destabilized eigenfunction $\hat{\Psi}_c$. {\bf b)} An elementary shear-transformation: the post-instability displacements that followed the instability of panel {\bf a)}.  {\bf c)} Nonaffine displacements $\vec{v}$ calculated at $\delta\gamma \equiv \gamma_c-\gamma = 0.007$. This delocalized field is used as the initial conditions $\hat{z}_{\mbox{\tiny ini}}$ for the minimization of $b(\hat{z})$ (see main text), the result of which is the plastic mode $\hat{\pi}$ displayed~in panel {\bf f)}. Panels {\bf d)} and {\bf e)} are intermediate states along the minimization of $b(\hat{z})$.}
\label{fig1}
\end{figure}

Plastic flow of disordered solids subjected to external loading is known to occur via localized rearrangements of small sets of particles, coined shear-transformations~\cite{argon_st}. Such rearrangements have been identified in experiments on bubble rafts \cite{bubble_raft}, foams \cite{dennin_foam_stzs}, emulsions \cite{pine_emulsions_stzs, weeks_emulsions_suspensions}, and colloidal glasses \cite{weeks_emulsions_suspensions,schall_stz_colloids}, as well as in atomistic computer simulations of model glasses \cite{argon_simulations,falk_langer_stz}. An example of such a shear-transformation, observed in a model glass in two dimensions deformed under athermal, quasi-static (AQS) shear \cite{footnote1}, is displayed in Fig.~\ref{fig1}b. Shear-transformations are known to self-organize in spatially correlated patterns \cite{lemaitre2004_avalanches, barrat_yield,lemaitre2006_avalanches,lemaitre_strain_rate,my_avalanches,salerno_robbins,lemaitre_role_of_temperature} in solids subjected to large stresses and low deformation rates. Their densities and other statistical properties, and mechanical consequences, are a subject of much recent debate~\cite{exist,eran_STZ,Vandembroucq,Bocquet,Talamali,jie1,jie2,jie3,bocquet_continuum,barrat_consequence}. Two questions, central to theoretical descriptions of elasto-plasticity, that we address in this work, are: can shear-transformations be predicted a-priori, and, if so, how?

The micro-mechanical process in which an athermal disordered solid destabilizes under quasi-static deformation is understood, asymptotically close to an instability strain $\gamma_c$, as a saddle-node bifurcation of the potential energy $U$ \cite{lemaitre2006_avalanches,lemaitre2004}. The immediate precursors to shear-transformations at strains $\gamma \to \gamma_c$ are identified as destabilized eigenfunctions $\hat{\Psi}_c$ (i.e.~their associated eigenvalues vanish at $\gamma_c$) of the dynamical matrix ${\cal M}_{ij} = \frac{\partial^2U}{\partial\vec{x}_i\partial\vec{x}_j}$, where $\vec{x}_i$ denotes the coordinate vector of the $i^{\mbox{\tiny th}}$ particle. Such an eigenfunction is presented in Fig.~\ref{fig1}a \cite{footnote1}. In the following, we refer to such eigenfunctions as \emph{destabilized modes}, to distinguish them from the post-instability displacements of particles -- agglomerations of shear-transformations -- which can be spatially extended. In Fig.~\ref{fig1}b we demonstrate that, when the post-instability displacements are not spatially extended, but rather form an isolated elementary shear transformation, their spatial structure is very similar to that of the destabilized mode. In contrast with the post-instability displacements that depend in general on a specific choice of dynamics~\cite{salerno_robbins}, and on external control parameters such as temperature~\cite{lemaitre_role_of_temperature}, strain~\cite{barrat_yield}, and strain rate \cite{lemaitre_strain_rate}, the spatial stucture of destabilized modes is an {\bf intrinsic characteristic} of the multi-dimensional potential energy function, and is therefore the focus of the present study.

A robust \emph{mechanical} definition of the precursors of plastic instabilities away from instability strains has not yet been put forward. Much effort has been dedicated recently to studying the role played by low-frequency normal modes in determining these precursors \cite{tanguy2010,vincenzo_epl_2010,manning2011,rottler_normal_modes,manning2015}. 
One key difficulty encountered in such studies is that low-frequency plane-waves, which have no appreciable effect on plasticity \cite{exist}, dominate the lower parts of the spectra of conventional model glasses, thus hindering attempts to use low-frequency modes to define flow-defect densities and correlate them with rates of plastic flow.

Another difficulty, which has been largely overlooked in the context of elasto-plasticity, is that mere frequencies of normal modes are not indicative of their relevance to plastic processes. In fact, modes which lead to mechanical instabilities (i.e.~take the system over energy barriers and into neighboring inherent states) appear as eigenfunctions of the dynamical matrix only very close to plastic instabilities, giving rise to difficulties in their detection and statistical quantification. Here we show that the effective detection of such modes away from plastic instabilities necessitates the consideration of the degree of \emph{asymmetry} associated to the variation $\delta U$ of the potential energy upon displacing the particles along those modes. We provide a theoretical framework that naturally embeds a mechanical definition of the precursors to plastic instabilities, and which effectively accounts for the said asymmetry. 

\section{theoretical framework}

We begin the discussion by considering an athermal elastic solid, of $N$ particles in $d$ dimensions, and let $\hat{z}$ denote a $Nd$ dimensional unit vector, i.e.~$\hat{z}_i\cdot\hat{z}_i =1$. Here and in what follows, repeated indices, labeling particles, are understood to be summed over, unless indicated otherwise. The coordinates $\vec{x}$ are displaced in the direction defined by $\hat{z}$ according to $\delta\vec{x} = s\hat{z}$, and we expand the potential energy $U$ as:
\begin{equation}\label{cubic}
\delta U_{\hat{z}}(s) \equiv U_{\hat{z}}(s) - U_0 \simeq \sFrac{1}{2}\kappa_{\hat{z}}s^2 + \sFrac{1}{6}\tau_{\hat{z}}s^3\,,
\end{equation}
where $U_0$ is the energy of the minimum in which the system resides, $\kappa_{\hat{z}}\!\equiv{\cal M}_{ij}:\!\hat{z}_i\hat{z}_j$ is the stiffness associated to $\hat{z}$, and $\tau_{\hat{z}}\!\equiv\!\frac{\partial^3U}{\partial \vec{x}_i\partial \vec{x}_j\partial \vec{x}_k}\!\tripleCdot\!\hat{z}_i\hat{z}_j\hat{z}_k$ is referred to in the following as the \emph{asymmetry} associated to $\hat{z}$. Within this cubic expansion, stationary points occur at $s=0$ and $s_\star(\hat{z}) = -2\frac{\kappa_{\hat{z}}}{\tau_{\hat{z}}}$; $s=0$ corresponds to the minimum in which the system resides, while $s_\star$ represents the saddle point (energy barrier) that separates this minimum and a neighboring inherent state. We thus define the energy difference between these stationary points, within the cubic expansion, as our \emph{barrier function}:
\begin{equation}\label{barrier_function}
b(\hat{z}) \equiv \sFrac{1}{2}\kappa_{\hat{z}}s_\star^2 + \sFrac{1}{6}\tau_{\hat{z}}s_\star^3 = \frac{2}{3}\frac{\kappa^3_{\hat{z}}}{\tau^2_{\hat{z}}}\,
\end{equation}

We emphasize that $b(\hat{z})$ is defined for a particular configuration of an elastic solid in mechanical equilibrium, and is a function of the direction $\hat{z}$. It has a rough landscape \cite{footnote4}; in this work we focus on directions $\hat{\pi}$ that correspond to local minima of $b(\hat{z})$. We refer to these directions in what follows as \emph{plastic modes}. From the definition of $b(\hat{z})$ it is clear that plastic modes $\hat{\pi}$ are associated with small stiffnesses $\kappa_{\hat{\pi}}$ and large asymmetries $\tau_{\hat{\pi}}$; they can be found numerically by minimizing $b(\hat{z})$ over directions $\hat{z}$, starting from some initial direction $\hat{z}_{\mbox{\tiny ini}}$, as demonstrated in panels {\bf c)}-{\bf f)} of Fig.~\ref{fig1}. Small $b(\hat{z})$'s should appropriately describe low saddle points (barriers) that separate the system from neighboring inherent states. We therefore expect modes $\hat{\pi}$ that correspond to low-lying minima of $b(\hat{z})$ (which are found by chosing an appropriate $\hat{z}_{\mbox{\tiny ini}}$ for the minimization), to encode information about imminent plastic instabilities. 

\begin{figure}[!ht]
\centering
\includegraphics[scale = 0.57]{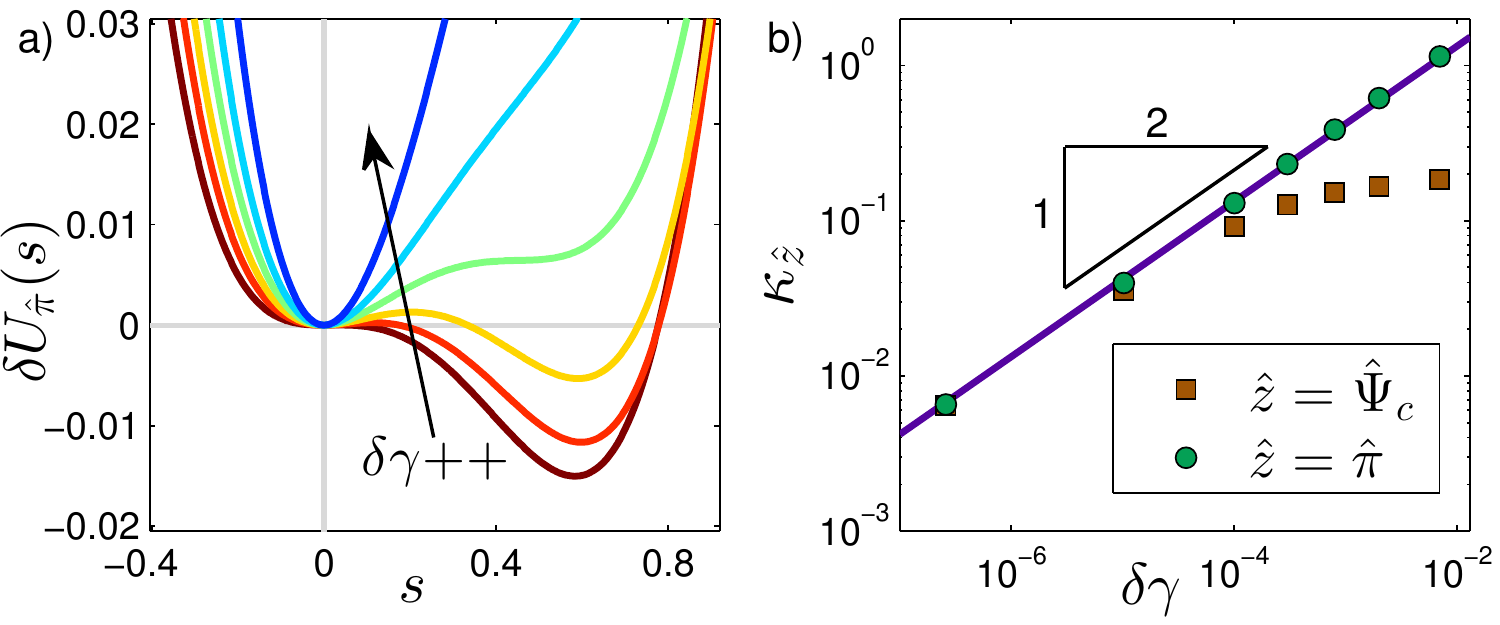}
\caption{\footnotesize (color online). {\bf a)} Variations $\delta U_{\hat{\pi}}(s)$ of the potential energy upon displacing the particles a distance $s$ along plastic modes $\hat{\pi}$, obtained as described in the text. The curves correspond to $\delta\gamma = 10^{-5}, 10^{-4}, 3\!\times\!10^{-4}, 8\!\times\!10^{-4}, 2\!\times\!10^{-3},7\!\times\!10^{-3}$. {\bf b)} Stiffnesses $\kappa_{\hat{z}} = {\cal M}_{ij}:\hat{z}_i\hat{z}_j$ associated to (circles) the plastic modes $\hat{\pi}$ used to calculate the variations plotted in panel {\bf a)}, and to (squares) the destabilized eigenfunction $\hat{\Psi}_c$, \emph{vs.}~$\delta \gamma$.}
\label{fig1b}
\end{figure}

\subsection{Numerical demonstration}

The approach described above is demonstrated in Fig.~\ref{fig1}; in panel {\bf a)} we display a destabilized mode $\hat{\Psi}_c$ calculated at the firstly-encountered plastic instability in an athermally sheared model glass, here at a strain $\gamma_c=0.011521$. Prior to this instability, at strains $\gamma = \gamma_c - \delta\gamma$, the nonaffine displacement field $\vec{v}_i \equiv -{\cal M}_{ij}^{-1}\cdot\frac{\partial^2U}{\partial \vec{x}_j\partial\gamma}$ is calculated \cite{lemaitre2004}. An example of $\vec{v}$, calculated at $\delta\gamma = 0.007$, is shown in panel {\bf c)}. At this distance (in strain) from the instability, the nonaffine displacements $\vec{v}$ are largely delocalized. We use the normalized $\hat{v} = \vec{v}/||\vec{v}||$ as the initial conditions $\hat{z}_{\mbox{\tiny ini}}$ for the minimization of $b(\hat{z})$; snapshots along the minimization are displayed in panels {\bf d)} and {\bf e)}. Upon convergence, we find a local minimum in the direction $\hat{\pi}$, which is displayed in panel {\bf f)}. The resemblance between $\hat{\pi}$ and the destabilized mode $\hat{\Psi}_c$ is striking: both the geometry and the core location appear to agree perfectly. 

This protocol is carried out over a broad range of intervals $\delta\gamma$, as specified in the caption of Fig.~\ref{fig1b}. For each $\delta\gamma$, after finding $\hat{\pi}$ as described above, we calculated its associated energy variation $\delta U_{\hat{\pi}}(s)$ and stiffness $\kappa_{\hat{\pi}}$, which are displayed in panels {\bf a)} and {\bf b)} of Fig.~\ref{fig1b}, respectively. In this example, already at a distance of the order $\delta \gamma \approx 10^{-3}$ to the instability strain, following the plastic mode ${\hat{\pi}}$ would carry the system above an energy barrier and into a neighboring minimum. 

\section{Destabilization of plastic modes}

We also plot in Fig.~\ref{fig1b}b the stiffness $\kappa_{\hat{\Psi}_c}$ associated to the destabilizing mode $\hat{\Psi}_c$. We find that only very close to the instability ($\delta\gamma \lesssim 10^{-5}$), the scaling $\kappa_{\hat{\Psi}_c}\!\sim\!\sqrt{\delta\gamma}$ holds \cite{footnote2}, whereas the stiffness associated to $\hat{\pi}$ follows $\kappa_{\hat{\pi}}\!\sim\! \sqrt{\delta\gamma}$ up to strains of order 1\% away from the instability. This finding supports the robustness of our definition of plastic modes, and the usefulness of our framework. It also supports the picture proposed by a number of recent studies \cite{Vandembroucq,Bocquet,Talamali,jie1,jie2,jie3}, that assumes the (reversible) destabilization process of a `soft spot' in a deformed glass is predominantly coupled to the external load, and not to other coexisting (reversible) destabilization processes. 

The scaling $\kappa_{\hat{\pi}}\!\!\sim\!\!\sqrt{\delta\gamma}$ can be derived as follows; modes ${\hat{\pi}}$ pertain to local minima of $b(\hat{z})$ and therefore satisfy $\frac{\partial b}{\partial\vec{z}}\big|_{\vec{z}= \hat{\pi}}\! =\! 0$, which implies that (see Appendix):
\begin{equation}\label{foo00}
\frac{\partial^3U}{\partial \vec{x}_i\partial \vec{x}_j\partial \vec{x}_k}\tripleCdot\hat{\pi}_j\hat{\pi}_k = \frac{\tau_{\hat{\pi}}}{\kappa_{\hat{\pi}}}{\cal M}_{ij}\cdot\hat{\pi}_j\,.
\end{equation}
Using this relation, we calculate the leading order variation of the stiffness $\kappa_{\hat{\pi}}$ with strain as
\begin{eqnarray}
\!\!\!\!\!\frac{d\kappa_{\hat{\pi}}}{d\gamma} & \simeq  & \frac{d{\cal M}_{ij}}{d\gamma}\!:\!\hat{\pi}_i\hat{\pi}_j \simeq \frac{\partial^3U}{\partial \vec{x}_i\partial \vec{x}_j\partial \vec{x}_k}\tripleCdot\hat{\pi}_i\hat{\pi}_j\vec{v}_k \nonumber \\
& = & -\frac{\tau_{\hat{\pi}}}{\kappa_{\hat{\pi}}}\hat{\pi}_i\!\cdot\!{\cal M}_{ij}\!\cdot\!{\cal M}^{-1}_{jk}\!\cdot\!\frac{\partial^2U}{\partial \vec{x}_k\partial\gamma} = -\frac{\tau_{\hat{\pi}}}{\kappa_{\hat{\pi}}}\hat{\pi}_i\!\cdot\!\frac{\partial^2U}{\partial \vec{x}_i\partial\gamma}.
\end{eqnarray}
As $\gamma \to \gamma_c$, $\kappa_{\hat{\pi}} \to 0$, but $\tau_{\hat{\pi}}\hat{\pi}_i\cdot\frac{\partial^2U}{\partial \vec{x}_i\partial\gamma}$ goes to a constant, yielding the differential scaling relation $\frac{d\kappa_{\hat{\pi}}}{d\gamma} \sim -\frac{1}{\kappa_{\hat{\pi}}}$, and thus the observed scaling $\kappa_{\hat{\pi}} \sim \sqrt{\delta\gamma}$. 

The resolution of the plastic mode as seen in Fig.~\ref{fig1} uses the nonaffine displacements $\vec{v}$ as the heuristic guess for $\hat{z}_{\mbox{\tiny ini}}$; this choice is made to demonstrate the usefulness of the framework -- despite the extended character of $\vec{v}$, it has a large overlap with the plastic mode $\hat{\pi}$, and thus resides in the basin of $\hat{\pi}$ on the landscape of $b(\hat{z})$. Obtaining the full field of plastic modes, however, requires using other heuristic $\hat{z}_{\mbox{\tiny ini}}$'s, that reside in basins that belong to other plastic modes. We leave the investigation of the optimal heuristics for the detection of the full field of plastic modes for future work.


\begin{figure}[ht]
\centering
\includegraphics[scale = 0.52]{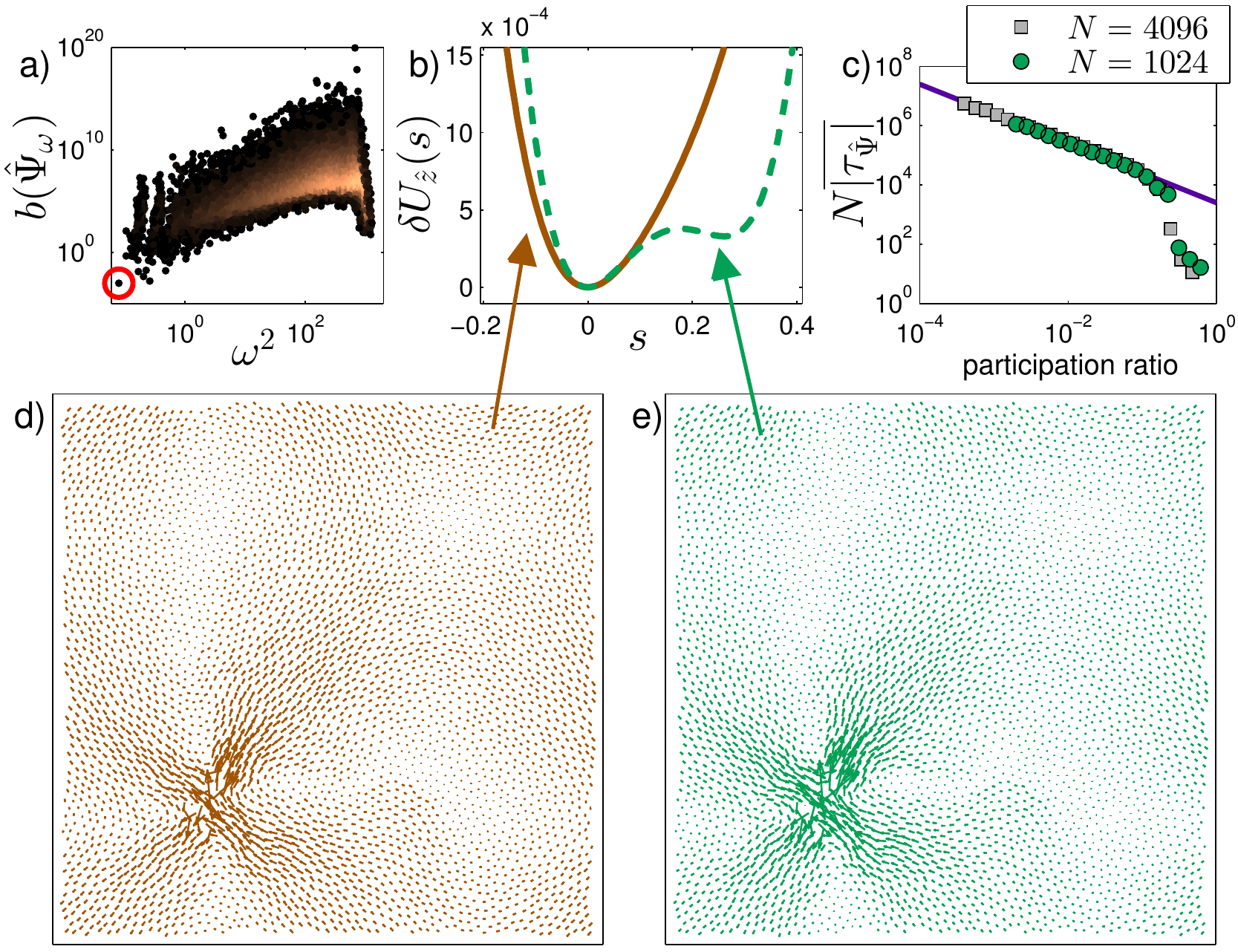}
\caption{\footnotesize (color online).  {\bf a)} Scatter plot of the barrier function Eq.~(\ref{barrier_function}) evaluated for eigenfunctions $\hat{\Psi}_\omega$ of ${\cal M}$, \emph{vs}.~their eigenvalues $\omega^2$. The eigenfunction $\hat{\Psi}_{\mbox{\tiny min}}$ represented by the circled data point is plotted in panel {\bf d)}, and used as the initial conditions $\hat{z}_{\mbox{\tiny ini}}$ for the minimization of the barrier function $b(\hat{z})$; the resulting plastic mode $\hat{\pi}$ is displayed in panel {\bf e)}. {\bf b)} Variations $\delta U_{\hat{z}}(s)$, calculated by displacing the particles according to $\delta\vec{x} = s\hat{\Psi}_{\mbox{\tiny min}}$ (continuous curve) and by $\delta\vec{x} = s\hat{\pi}$ (dashed curve).  {\bf c)} The products $N|\tau_{\hat{\Psi}}|$, averaged over bins of the participation ratio, see text for definitions. }
\label{fig2}
\end{figure}

\section{Comparison to normal modes}

How indicative are normal modes of imminent plastic instabilities, compared to plastic modes? In Fig.~\ref{fig2}a we present a scatter-plot of the barrier function evaluated at normal modes $\hat{\Psi}_\omega$, \emph{vs}.~the square of their associated frequencies $\omega^2$, calculated for a few tens of undeformed (isotropic) solid realizations. A clear trend appears: smaller values of $b(\hat{\Psi}_\omega)$ are found for lower-frequency modes. The circled data point represents the mode $\hat{\Psi}_{\mbox{\tiny min}}$ associated to the lowest value of $b(\hat{\Psi}_\omega)$ amongst all modes calculated; it is displayed in panel {\bf d)}. Remarkably, this normal mode displays the same spatial features as observed for destabilized modes, reinforcing that $b(\hat{z})$ is indeed sensitive to `plastic-like' modes. The variation $\delta U_{\hat{\Psi}_{\mbox{\tiny min}}}(s)$ is plotted in panel {\bf b)} (continuous line). Despite possessing the smallest $b$ amongst our entire ensemble of modes, $\delta U_{\hat{\Psi}_{\mbox{\tiny min}}}(s)$ displays only a slight asymmetry between positive and negative displacements $s$, and the energy monotonically increases with $|s|$. Using $\hat{\Psi}_{\mbox{\tiny min}}$ as the initial condition $\hat{z}_{\mbox{\tiny ini}}$ for the minimization of $b(\hat{z})$, we find the plastic mode $\hat{\pi}$ displayed in panel {\bf e)}. On the face of it, $\hat{\Psi}_{\mbox{\tiny min}}$ and $\hat{\pi}$ appear to be very similar in their spatial structure and geometry. However, examining the corresponding variation $\delta U_{\hat{\pi}}(s)$, represented by the dashed line in panel {\bf b)}, reveals a dramatic difference between them: following~$\hat{\pi}$~takes the system over a energy barrier, to a neighboring minimum.

We further utilize our ensemble of normal modes to study the relation between the degree of localization of modes and their associated asymmetries $\tau_{\hat{\Psi}}$. A similar analysis was carried out in \cite{vincenzo_epl_2010} in the context~of the unjamming point \cite{ohern2003,liu_review,van_hecke}. We quantify the degree of localization of a mode $\hat{\Psi}$ via its participation ratio $e\!=\![N\sum_i (\hat{\Psi}_i\cdot\hat{\Psi}_i)^2]^{-1}$; localized modes have $e\!\sim\! N^{-1}$, whereas maximally delocalized modes have $e\!\sim\!1$. In Fig.~\ref{fig2}c we plot the means $\overline{|\tau_{\hat{\Psi}}|}$ \cite{footnote3}, averaged over modes $\hat{\Psi}$ with similar participation ratios, for systems of $N\!=\!1024$ and $N\!=\!4096$. We find that for participation ratios $e<10^{-1}$, the asymmetries follow $\overline{|\tau_{\hat{\Psi}}|}\!\sim\! (eN)^{-1}$. This can be explained with the following simple model: if there are effectively $N^\alpha$ non-zero components in a normal mode ($0\!<\!\alpha\!<\!1$), normalization then requires that a characteristic non-zero component is of magnitude $||\hat{\Psi}_i||\!\sim\! N^{-\frac{\alpha}{2}}$. The participation ratio is then expected to follow $e\!\sim\! N^{\alpha-1}$ (due to summing over positive terms). Since the pairwise potential is short ranged, and the tensor elements $\frac{\partial^3U}{\partial \vec{x}_i\partial \vec{x}_j\partial \vec{x}_k}$ are of either sign, then $\tau_{\hat{\Psi}}$ consist of a sum over $N^\alpha$ terms, each of order $||\hat{\Psi}_i||^3\!\sim\! N^{-\frac{3\alpha}{2}}$, of random signs, and we therefore expect $\tau_{\hat{\Psi}}\!\sim N^{-\alpha}\!\sim (eN)^{-1}$, in consistency with our measurement. For participation ratios $e>10^{-1}$, this relation breaks down, and asymmetries are much \emph{smaller} than what is predicted by this simple model, which assumes that normal modes are random objects. Nevertheless, the same trend remains unchanged: delocalized modes are associated, on average, with more symmetric variations of the energy. These observations explain the localized nature of plastic instabilities found in deformed glasses, as can be seen, e.g., in Fig.~\ref{fig1}a.

\begin{figure}[ht]
\centering
\includegraphics[scale = 0.52]{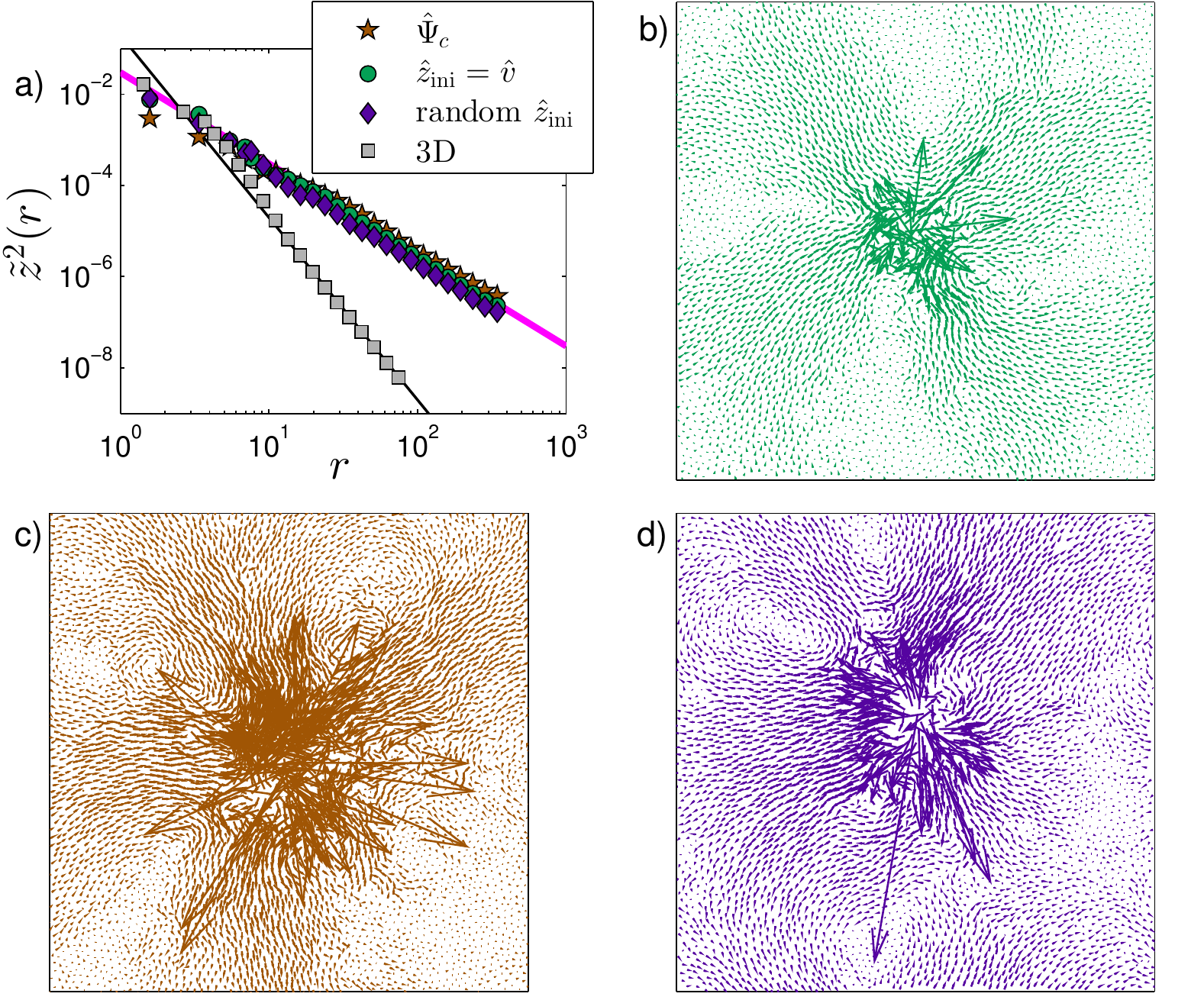}
\caption{\footnotesize (color online). {\bf a)} Spatial decay of plastic modes, see text for definition. All modes analyzed decay as $r^{1-d}$, as indicated by the continuous lines, where $r$ is the distance to the core center. {\bf b)} A plastic mode found by chosing a random $\hat{z}_{\mbox{\tiny ini}}$. {\bf c)} A plastic mode calculated in a disordered network of relaxed Hookean springs. {\bf d)} A plastic mode found in a Lennard-Jones glass under isotropic tension.}
\label{fig3}
\end{figure}

\section{Structure of plastic modes}

To characterize the spatial structure of plastic modes, we define $\tilde{z}^2(r)$ as the median of the squared magnitude of the components $\hat{z}_i\cdot\hat{z}_i$ (no summation implied), taken over a shell of thickness on the order of the nearest-neighbor distance, and of radius $r$ away from the core of the plastic mode (detecting the locus of the core is explained in the Appendix). In Fig.~\ref{fig3}a we compare the spatial decay of two plastic modes, one obtained by setting $\hat{z}_{\mbox{\tiny ini}}$ to be the direction of the nonaffine displacements (see definition above and Fig.~\ref{fig1}), and the other by setting $\hat{z}_{\mbox{\tiny ini}}$ to be a random direction. These decay profiles are also compared to that of a destabilized mode $\hat{\Psi}_c$. We also show the decay profile of a plastic mode calculate in a 3D solid. We find that at distances $r$ away from the core, plastic modes decay as $r^{1-d}$. Remarkably, this is the same decay law found for the linear responses of displacements to dipolar point forces \cite{bocquet_continuum,breakdown}. 

In Fig.~\ref{fig3}b we present a plastic mode obtained with a random $\hat{z}_{\mbox{\tiny ini}}$. We find that this mode shares the same geometric features as the destabilized modes $\hat{\Psi}_c$ upon shear-induced plastic instabilities -- a disordered core, and a long-ranged affine quadrupolar shear-like displacement field away from the core \cite{lemaitre2006_avalanches,lemaitre2004,bocquet_continuum}. We thus conclude that plastic modes $\hat{\pi}$ associated to different local minima of $b(\hat{z})$ share similar structrural features, that do not depend on the particular minima to which they correspond.

\subsection{Effects of loading conditions}

We finally examine how the geometry of plastic modes depends on the loading conditions imposed on the solid. In panels {\bf c)} and {\bf d)} of Fig.~\ref{fig3}, two additional examples of plastic modes $\hat{\pi}$ obtained from a random $\hat{z}_{\mbox{\tiny ini}}$ are displayed; $\hat{\pi}$ of panel {\bf c)} was calculated in a disordered network of relaxed Hookian springs (all springs are neither stretched nor compressed) with an average of 4.1 springs connected to each node. It displays a similar spatial structure as that of plastic modes found in model glasses that are prestressed, i.e.~in which finite forces are exerted between the constituent particles \cite{prestress_footnote}. Our findings indicate that proximity to prestress-induced micro-mechanical buckling instabilities \cite{matthieu_PRE_2005} is not the origin of the generic structure of plastic modes.

The plastic mode $\hat{\pi}$ of panel {\bf d)} of Fig.~\ref{fig3} was calculated in a Lennard-Jones glass (with a pairwise potential that includes an attractive term, see Appendix for details) under isotropic tension, just before macroscopic failure (here $-p/B \approx 10^{-2}$ is at least $80\%$ of the yield strain, where $p$ is the pressure and $B$ is the bulk modulus). We find in this case that in addition to the clear shear-like displacements that are typically seen in plastic modes found in glasses under compressive stresses, the dilatant part of the displacements due to the tensile loading conditions is apparent. We conclude that the loading conditions imposed on a solid can be reflected in the geometric features of its plastic modes; we leave the systematic study of this dependence for future work.

\section{Summary and discussion}

In this work we demonstrated that modes $\hat{\pi}$ corresponding to local minima of the barrier function $b(\hat{z})$, coined \emph{plastic modes}, are indicative of directions in configuration space that lead to plastic instabilities, and more so compared to the most localized low-frequency normal modes. As such, our approach can serve as a solid basis for instability-detection algorithms. Such algorithms are highly desirable, as they can put to test theoretical frameworks of elasto-plasticity that involve the dynamics of a population of `soft-spots'. These algorithms need not be restricted to the investigation of plastic flow in disordered solids; the generality of our framework would render them suitable for studying a diverse set of systems, including dislocated crystalline solids, deeply supercooled liquids and proteins.

Furthermore, our theoretical framework explains the origin of the localized nature of plastic instabilities. Building on our framework, we predict that the stiffness associated to plastic modes follows $\kappa \sim \sqrt{\gamma_c - \gamma}$, and show numerically that this scaling holds over a large range of strains away from an instability strain $\gamma_c$. This adds relevance to recently proposed models that assume reversible destabilization processes of soft spots are decoupled from each other. Finally, we have investigated the spatial features of plastic modes, and provide evidence that the detailed geometry of plastic modes is sensitive to the loading conditions imposed on the solid.

\begin{acknowledgments}
We thank Eran Bouchbinder, Smarajit Karmakar, Gustavo D\"uring, Itamar Procaccia and Matthieu Wyart for fruitful discussions. 
\end{acknowledgments}

\appendix
\section{Models and numerical methods}

\subsection{Calculation of plastic modes}

We begin with describing our approach to calculating plastic modes; starting from an initial guess $\hat{z}_{\mbox{\tiny ini}}$, we minimize the barrier function:
\begin{equation}\label{foo00}
b(\hat{z}) \equiv \frac{2}{3} \frac{\left(\frac{\partial^2U}{\partial\vec{x}_i\partial\vec{x}_j}:\hat{z}_i\hat{z}_j\right)^3}
{\left(\frac{\partial^3U}{\partial\vec{x}_i\partial\vec{x}_j\partial\vec{x}_k}\tripleCdot\hat{z}_i\hat{z}_j\hat{z}_k\right)^2}\,,
\end{equation}
over directions $\hat{z}$ in the $Nd$-dimensional configuration space. Here $U(\vec{x})$ is the potential energy, $\vec{x}_i$ denotes the Cartesian coordinate vector of the $i^{\mbox{\tiny th}}$ particle, and repeated indices are understood to be summed over. Minimizations are carried out using a standard nonlinear conjugate gradient minimization algorithm, which was modified such that the norm $||\hat{z}||$ remains of order unity, to maintain numerical stability. Notice the invariance $b(\hat{z})=b(s\hat{z})$ for any finite length $s$, which allows us to control the norm without effecting the minimization process. 

Panels {\bf c)}-{\bf f)} of Fig.~1 in the main text show an example of a minimization of the barrier function. Panel {\bf c)} is the initial guess $\hat{z}_{\mbox{\tiny ini}} = \hat{v}$ where $\vec{v}$ denotes the nonaffine displacement field, as defined in the main text. Panels {\bf d)} and {\bf e)} displays states encountered along the minimization algorithm, and panel {\bf f)} is the plastic mode obtained upon convergence of the minimization algorithm to a local minimum of $b(\hat{z})$.


\subsection{Mode spatial structure analysis}

The analyses of the spatial decay of modes $\hat{z}$ were carried out as follows. First, we find the center of the core $\vec{x}_{\mbox{\tiny center}}$, which we determine following 
\begin{equation}
\vec{x}_{\mbox{\tiny center}}= \frac{\sum_{i\in\Omega} ||\hat{z}_i||^2 \vec{x}_i}{\sum_{i\in\Omega} ||\hat{z}_i||^2},
\end{equation}
where $\Omega$ denotes the set of the $w$ particles with the highest $||\hat{z}_i||^2$. The choice $w=4$ was found to be the optimal choice for most of the plastic modes we analyzed, and we never used $w>10$. We then divided the area of the system into rings (or shells in the 3D case) of radius $r$ away from the center. The thickness of the rings was set such that there were at least 10 particles in each ring. For each ring, we found $\tilde{z}^2(r)$ defined as the {\bf median} value over all $||\hat{z}_i||^2$ in that ring. 

\subsection{Model definitions}

We continue with describing the models used in this work, starting with the model glass for which most data are presented: a 50:50 binary mixture of `large' and `small' particles of equal mass $m$, interacting via radially-symmetric purely repulsive inverse power-law pairwise potentials, that follow
\begin{equation}
\varphi_{\mbox{\tiny IPL}}(r_{ij}) = \left\{ \begin{array}{ccc}\varepsilon\left[ \left( \sFrac{\lambda_{ij}}{r_{ij}} \right)^n + \sum\limits_{\ell=0}^q c_{2\ell}\left(\sFrac{r_{ij}}{\lambda_{ij}}\right)^{2\ell}\right]&,&\sFrac{r_{ij}}{\lambda_{ij}}\le x_c\\0&,&\sFrac{r_{ij}}{\lambda_{ij}}> x_c\end{array} \right.,
\end{equation}
where $r_{ij}$ is the distance between the $i^{\mbox{\tiny th}}$ and $j^{\mbox{\tiny th}}$ particles, $\varepsilon$ is an energy scale, and $x_c$ is the dimensionless distance for which $\varphi_{\mbox{\tiny IPL}}$ vanishes continuously up to $q$ derivatives. Distances are measured in terms of the interaction lengthscale $\lambda$ between two `small' particles, and the rest are chosen to be $\lambda_{ij} = 1.18\lambda$ for one `small' and one `large' particle, and $\lambda_{ij} = 1.4\lambda$ for two `large' particles. The coefficients $c_{2\ell}$ are given by
\begin{equation}
c_{2\ell} = \frac{(-1)^{\ell+1}}{(2q-2\ell)!!(2\ell)!!}\frac{(n+2q)!!}{(n-2)!!(n+2\ell)}x_c^{-(n+2\ell)}\,.
\end{equation}
We chose the parameters $x_c = 1.48, n=10$, and $q=3$. The density was set to be $N/V = 0.86\lambda^{-2}$ for our 2D systems, and $N/V = 0.82\lambda^{-3}$ for our 3D systems. We note that with these densities two coordination shells fall within the interaction range. Both the 2D and 3D systems undergo a computer-glass-transition at temperatures of about $T_g\approx 0.5\varepsilon/k_B$. Solids were created by a fast quench from the melt to a target temperature $T\ll T_g$, followed by an energy minimization using a standard nonlinear conjugate gradient algorithm. We employed two-dimensional systems of $N=4096$ for producing the data presented in figures 1, 2, 3, and 4b of the main text. We measured decay profiles of plastic modes (plotted in Fig.~4a of the main text) in 2D systems of $N=409600$, and in 3D systems of $N=4096000$. Normal mode analyses were carried out using the linear algebra package LAPACK \cite{lapack}, and the numerical analysis software MATLAB \cite{matlab}. Systems were deformed by imposing a simple shear, namely each particle was displaced according to 
\begin{eqnarray}
x_i & \to & x_i + \delta\gamma y_i\,, \\
y_i & \to & y_i\,,
\end{eqnarray}
where $x_i,y_i$ denote the $x$ and $y$ coordinates of the $i^{\mbox{\tiny th}}$ particle. The strain increments $\delta\gamma$ used were not larger than $10^{-5}$. 

\begin{figure}
\centering
\subfigure[]{\includegraphics[width=6cm]{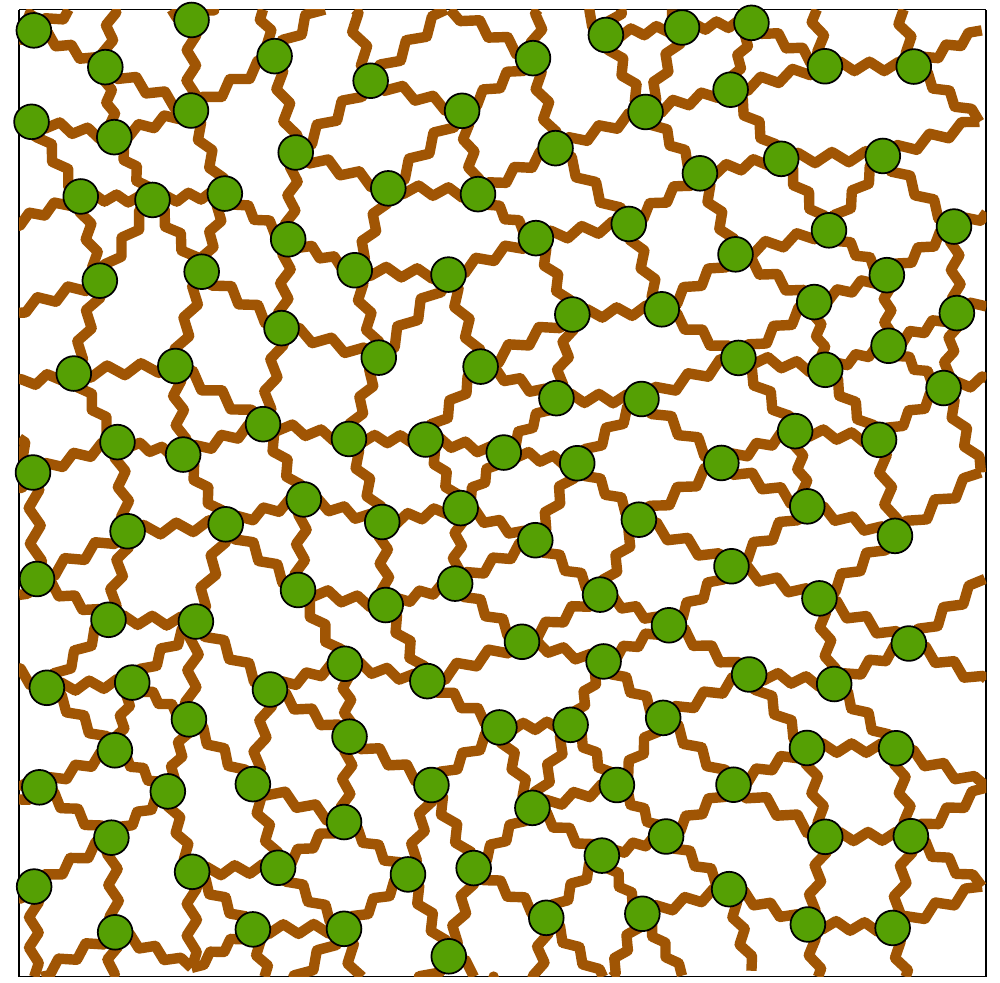}}
\hspace{1.5cm}
\subfigure[]{\includegraphics[width=6cm]{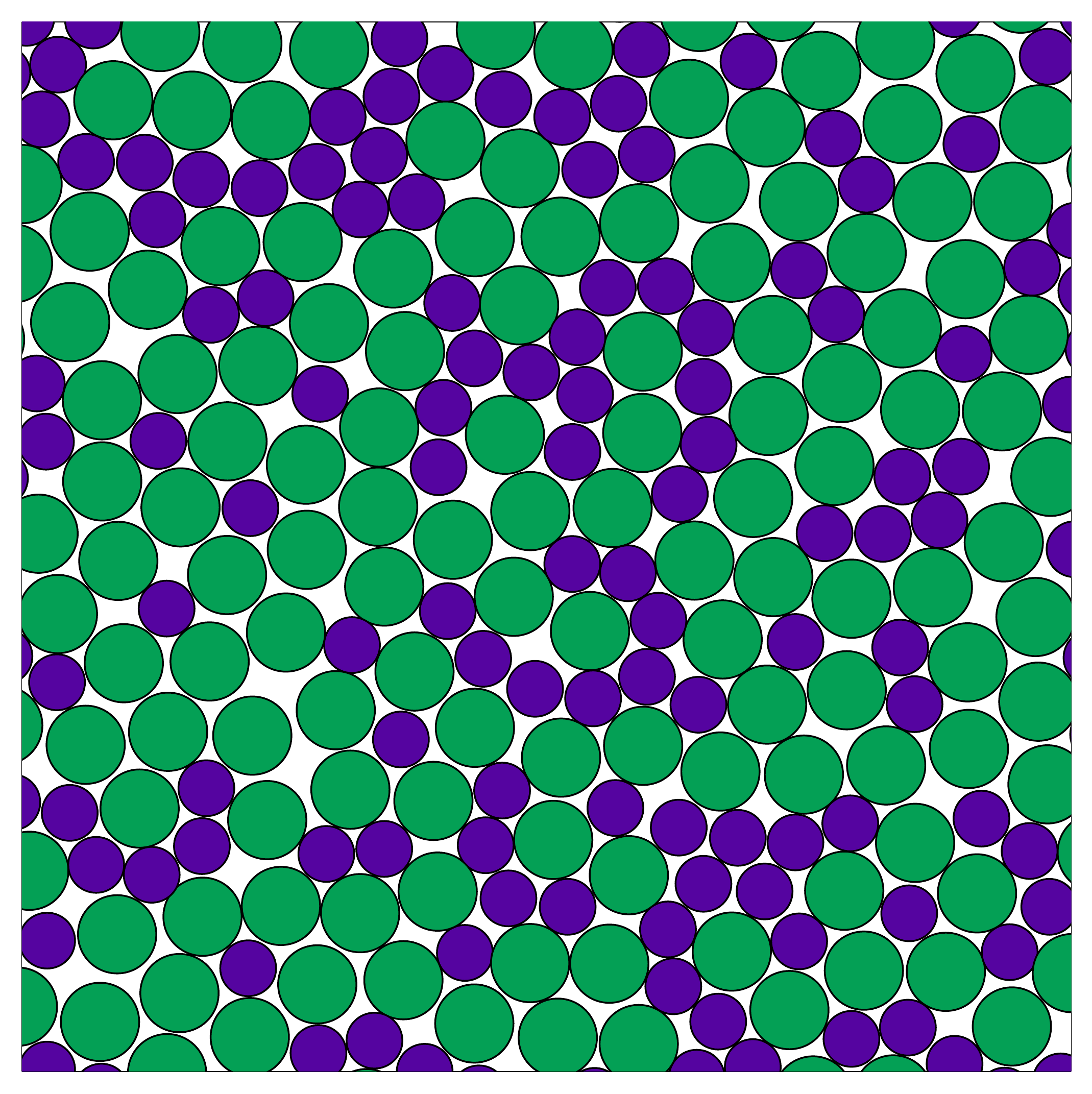}}
\caption{\footnotesize (color online). (a) An example of a two-dimensional disordered network of relaxed Hookean springs with a mean connectivity $\langle z \rangle \approx 4.1$, as used in this work. (b)  An example of a pre-failure two-dimensional Lennard-Jones solid.}
\label{foo_fig}
\end{figure}


\begin{widetext}

Plastic modes were also calculated in two more systems: 2D disordered networks of relaxed Hookean springs (see panel (a) of Fig.~\ref{foo_fig} in this SM), which were created as described in \cite{breakdown}, and in 2D, 50:50 binary mixtures of particles interacting via the Lennard-Jones potential (see Fig.~4 of main text):
\begin{equation}
\varphi_{\mbox{\tiny LJ}}(r_{ij}) = \left\{ \begin{array}{ccc}\varepsilon\left[ \left( \sFrac{\lambda_{ij}}{r_{ij}} \right)^{12} - \left( \sFrac{\lambda_{ij}}{r_{ij}} \right)^6 +c_4\left(\sFrac{r_{ij}}{\lambda_{ij}}\right)^4 + c_2\left(\sFrac{r_{ij}}{\lambda_{ij}}\right)^2 + c_0\right]&,&\sFrac{r_{ij}}{\lambda_{ij}}\le x_c\\0&,&\sFrac{r_{ij}}{\lambda_{ij}}> x_c\end{array} \right.\,.
\end{equation}
Here $x_c = 2.5$ is the dimensionless distance for which $\varphi_{\mbox{\tiny LJ}}$ vanishes continuously up to two derivatives, which implies that $c_4=0.0006201261686784$, $c_2=-0.00970155098112$, and $c_0 = 0.040490237952$. The interaction lengths $\lambda_{ij}$ are identical to those chosen for the inverse-power-law system described above, as is the choice of units of length $\lambda$, and $\varepsilon$ is an energy scale. Systems were first created at the initial density of $N/V=0.615\lambda^{-2}$, which resulted in a small but positive pressure of $p\approx0.05\varepsilon/\lambda^2$. We then applied small expansive strain increments $\delta\gamma$, by applying the transformation $\vec{x}_i \to (1+\delta\gamma)\vec{x}_i$ to the coordinates $\vec{x}_i$, and subsequently minimizing the energy. We find that the system fails via cavitation at tensile pressures of $p_{\mbox{\tiny fail}}\approx -0.22\varepsilon/\lambda^2$. The bulk modulus \cite{elasticity} in the pre-failure states is $B \approx 18\varepsilon/\lambda^2$, which translates to a yield strain of $p_{\mbox{\tiny fail}}/B \approx 10^{-2}$. An example of a pre-failure solid is displayed in panel (b) of Fig.~\ref{foo_fig} in this SM. 

\end{widetext}

\section{Shear coupling of plastic modes' stiffnesses}
In this final section we provide an outline of the formalism leading to Eq.~(4) of the main text, for $\frac{d\kappa_{\hat{\pi}}}{d\gamma}$. A detailed derivation will be presented elsewhere \cite{plastic_mode_coupling_paper}. We begin with noting that, by construction, the barrier function is invariant to inflations $b(\hat{z}) = b(s\hat{z})$, for any finite length $s$. This allows to define $b(\vec{z})$ to be a function of the independent variables $\vec{z}_i$, which is the form assumed below. 

For the sake of brevity we denote ${\cal M}_{ij}\equiv \frac{\partial^2U}{\partial \vec{x}_i\partial \vec{x}_j}$ and $U'''_{ijk} \equiv \frac{\partial^3U}{\partial \vec{x}_i\partial \vec{x}_j\partial \vec{x}_k}$, and write the gradient of $b(\vec{z})$ as
\begin{equation}
\frac{\partial b}{\partial \vec{z}_i} = 4\frac{\kappa_{\vec{z}}^2}{\tau_{\vec{z}}^2}\left( {\cal M}_{ij}\cdot\vec{z}_j - \frac{\kappa_{\vec{z}}}{\tau_{\vec{z}}}U'''_{ijk}:\vec{z}_j\vec{z}_k\right)\,.
\end{equation}
Modes $\hat{\pi}$ that correspond to local minima of $b$ satisfy $\frac{\partial b}{\partial \vec{z}_i}\big|_{\vec{z} = \hat{\pi}} = 0$, which immediately yields Eq.~(3) of the main text:
\begin{equation}\label{foo03}
U'''_{ijk}:\hat{\pi}_j\hat{\pi}_k = \frac{\tau_{\hat{\pi}}}{\kappa_{\hat{\pi}}}{\cal M}_{ij}\cdot\hat{\pi}_j\,.
\end{equation}
We next turn to calculating the total derivative with respect to strain $\gamma$ of the stiffness $\kappa_{\hat{\pi}}$ as
\begin{equation}\label{foo05}
\frac{d\kappa_{\hat{\pi}}}{d\gamma} = \frac{d{\cal M}_{ij}}{d\gamma}:\hat{\pi}_i\hat{\pi}_j + 2{\cal M}_{ij}:\hat{\pi}_i\frac{d\hat{\pi}_j}{d\gamma}\,.
\end{equation}
An equation for $\frac{d\hat{\pi}}{d\gamma}$ is obtained by requiring that $\hat{\pi}$ remains a local minimum of $b$ under the deformation, namely
\begin{equation}\label{foo02}
\frac{d}{d\gamma}\bigg|_{\vec{z}=\hat{\pi}}\frac{\partial b}{\partial\vec{z}_i} = 0\,.
\end{equation}
Using that
\begin{equation}
\frac{\partial^2b}{\partial \vec{z}_i\partial \vec{z}_j}\bigg|_{\vec{z}=\hat{\pi}}\cdot\hat{\pi}_j = \frac{\partial^2b}{\partial \vec{x}_i\partial \vec{z}_j}\bigg|_{\vec{z}=\hat{\pi}}\cdot\hat{\pi}_j = \frac{\partial^2b}{\partial\gamma\partial\vec{z}_i}\bigg|_{\vec{z}=\hat{\pi}}\cdot\hat{\pi}_i =  0\,,
\end{equation}
one finds that $(i)$ $\frac{d\hat{\pi}}{d\gamma}\cdot\hat{\pi}= 0$, and that $(ii)$ $||\frac{d\hat{\pi}}{d\gamma}||$ goes to a constant at the instability strain $\gamma_c$. However, $\frac{d{\cal M}}{d\gamma}$ is singular, and we are thus left with 
\begin{equation}
\frac{d\kappa_{\hat{\pi}}}{d\gamma} \simeq \frac{d{\cal M}_{ij}}{d\gamma}:\hat{\pi}_i\hat{\pi}_j\,,
\end{equation}
as seen in Eq.~(4) of the main text. Finally, following \cite{elasticity} and references within, the total derivative with respect to strain of the dynamical matrix is
\begin{equation}
\frac{d {\cal M}_{ij}}{d\gamma} = \frac{\partial {\cal M}_{ij}}{\partial \gamma} + \frac{\partial {\cal M}_{ij}}{\partial\vec{x}}\cdot\frac{d\vec{x}_k}{d\gamma}\,,
\end{equation}
where $\frac{d\vec{x}_i}{d\gamma} \equiv -{\cal M}^{-1}_{ij}\cdot\frac{\partial^2U}{\partial\vec{x}_j\partial\gamma}$ are the nonaffine displacements, denoted $\vec{v}$ in the main text. Since as $\gamma_c$ is approached, $||\vec{v}|| \sim (\gamma_c - \gamma)^{-\frac{1}{2}}$ \cite{elasticity}, to leading order we find
\begin{equation}
\frac{d {\cal M}_{ij}}{d\gamma} \simeq U'''_{ijk}\cdot\vec{v}_k\,,
\end{equation}
as seen in Eq.~(4) of the main text.


\begin{thebibliography}{9}
\bibitem{footnote4}
Finding its global minimum is desirable, but is currently a formidable task; we leave the design of algorithms that can find the global minimum of $b(\hat{z})$ for future studies.
\bibitem{footnote2}
In a system of linear size $L$, the scaling law $\kappa_{\hat{\Psi}}\sim\sqrt{\delta\gamma}$ is only expected to hold below $\delta\gamma \sim L^{-4}$ due to hybridizations with low-frequency plane waves.
\bibitem{footnote3}
We consider the absolute magnitudes $|\tau_{\hat{\Psi}}|$ since normal modes $\hat{\Psi}$, and therefore also their associated asymmetries $\tau_{\hat{\Psi}}$, are defined up to a sign.
\end{thebibliography}
\end{document}